# Application of the Multi-Peaked Analytically Extended Function to Representation of Some Measured Lightning Currents

**Karl Lundengård [1], Milica Rančić [1], Vesna Javor [2], Sergei Silvestrov [1]**

**Abstract:** A multi-peaked form of the analytically extended function (AEF) is used for approximation of lightning current waveforms in this paper. The AEF function's parameters are estimated using the Marquardt least-squares method (MLSM), and the general procedure for fitting the *p*-peaked AEF function to a waveform with an arbitrary (finite) number of peaks is briefly described. This framework is used for obtaining parameters of 2-peaked waveforms typically present when measuring first negative stroke currents. Advantages, disadvantages and possible improvements of the approach are also discussed.

## 1 Introduction

Many different types of systems, objects and equipment are susceptible to damage from lightning discharge. Lightning effects are usually analysed using lightning discharge models. Most of these models imply channel-base current functions. Various single and multi-peaked functions have been proposed in the literature, e.g. [4]-[10], [14]-[15]. For engineering and electromagnetic models, a general function that would be able to reproduce desired waveshapes is needed, such that analytical solutions for its derivatives, integrals, and integral transformations exist. A multi-peaked channel-base current function has been proposed in [8] as a generalization of the so-called TRF (two-rise front) function from [9], which possesses such properties.

In this paper we explore the fitting of the proposed multi-peaked function [14], [15], the so-called *p*-peaked analytically extended function (AEF), to either measured data, or other proposed functions [4], [6] able to produce a double-peaked waveshape. Explicit expressions for a number of basic AEF properties, such as derivatives and integrals, are given in [14], [15].

---

[1] Mälardalen University, Division of Applied Mathematics, UKK, Högskoleplan 1, Box 883, 721 23 Västerås, Sweden; E-mail: {karl.lundengard, milica.rancic, sergei.silvestrov}@mdh.se

[2] University of Niš, Dept. of Power Engineering, Faculty of Electronic Engineering, Aleksandra Medvedeva 14, 18000 Niš, Serbia: E-mail: vesna.javor@elfak.ni.ac.rs

Section 3 discusses fitting of the *p*-peaked AEF applying a general scheme that employs the Marquardt least-squares method (MLSM), [16]. Application of this framework to five different 2-peaked waveforms is illustrated in Section 4 through a number of numerical experiments. For validation, obtained waveforms are compared to either measured data, [2], [3], [17], or that simulated by other previously proposed functions, [4], [6].

Finally, some findings of performed analysis, and possible paths of future work are discussed in the conclusion.

## 2 The *p*-Peaked AEF

The elementary function used to construct the *p*-peaked AEF is given by

$$x(\beta;t)=\left(te^{1-t}\right)^{\beta},0\le t, \tag{1}$$

and is in [15] referred to as the *power exponential function*. Qualitatively, it is similar to desired waveshapes in the sense that its rising initial part is steep, and is followed by a very slowly decaying part. The $\beta$-parameter in (1) determines the steepness of both the rising and decaying part, which is illustrated in Fig. 1.

In order to get a function with multiple peaks and where the steepness of the rise between each peak as well as the slope of the decaying part is not dependent on each other, we define the analytically extended function (AEF) as a function that consist of piecewise linear combinations of the power exponential functions that have been scaled and translated so that the resulting function is continuous. Then, the *p*-peaked AEF current function is given by

$$i(t)=\begin{cases}\left(\sum_{k=1}^{q-1} I_{m_k}\right)+I_{m_q}\sum_{k=1}^{n_q}\eta_{q,k}x_{q,k}(t), & \begin{array}{c} t_{m_{q-1}}\le t\le t_{m_q}, \\ 1\le q\le p, \end{array} \\ \left(\sum_{k=1}^{p} I_{m_k}\right)\sum_{k=1}^{n_{p+1}}\eta_{p+1,k}x_{p+1,k}(t), & t_{m_p}\le t, \end{cases} \tag{2}$$

where:

- $I_{m_1}, I_{m_2}, \ldots, I_{m_p}$ - the difference in height between each pair of peaks,
- $t_{m_1}, \ldots, t_{m_p}$ - the times corresponding to these peaks,
- $n_q > 0$ - the number of terms in each interval; larger number gives more possible shapes but also adds parameters that need to be fitted,
- $\eta_{q,k}$ - real values s.t. the sum over $\eta_{q,k}$ for $1 \le k \le n_q$ is equal to unity,
- $x_{q,k}(t)$ - power exponential functions defined by $\beta_{q,k}$ parameters in the following way:

$$\begin{aligned} x_{q,k}(t) &= x\left(\beta_{q,k}^2+1;\frac{t-t_{m_{q-1}}}{t_{m_q}-t_{m_{q-1}}}\right), \quad 1\le q\le p, \\ x_{p+1,k}(t) &= x\left(\beta_{p+1,k}^2;\frac{t-t_{m_{q-1}}}{t_{m_q}-t_{m_{q-1}}}\right). \end{aligned} \tag{3}$$

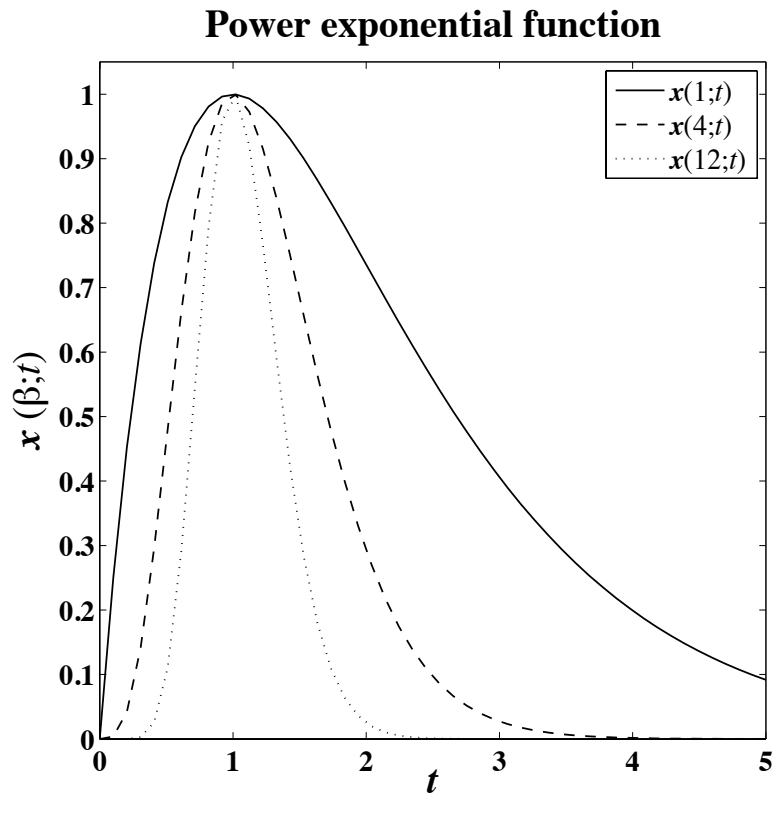

**Fig. 1 –** *Steepness dependence of the power exponential function on the $\beta$-parameter.*

The previous formulation of the $p$-peaked AEF function can be written more compactly as

$$i(t)=\begin{cases}\left(\sum_{k=1}^{q-1} I_{m_k}\right)+I_{m_q}\boldsymbol{\eta}_q^{\mathrm{T}}\mathbf{x}_q(t), & \begin{array}{c} t_{m_{q-1}}\le t\le t_{m_q}, \\ 1\le q\le p, \end{array} \\ \left(\sum_{k=1}^{p} I_{m_k}\right)\boldsymbol{\eta}_{p+1}^{\mathrm{T}}\mathbf{x}_{p+1}(t), & t_{m_p}\le t, \end{cases} \tag{4}$$

where we introduce the following vectors:

$$\begin{aligned} \boldsymbol{\eta}_q &= \begin{bmatrix} \eta_{q,1} & \eta_{q,2} & \cdots & \eta_{q,n_q} \end{bmatrix}^{\mathrm{T}}, \\ \mathbf{x}_q(t) &= \begin{bmatrix} x_{q,1}(t) & x_{q,2}(t) & \cdots & x_{q,n_q}(t) \end{bmatrix}. \end{aligned}$$

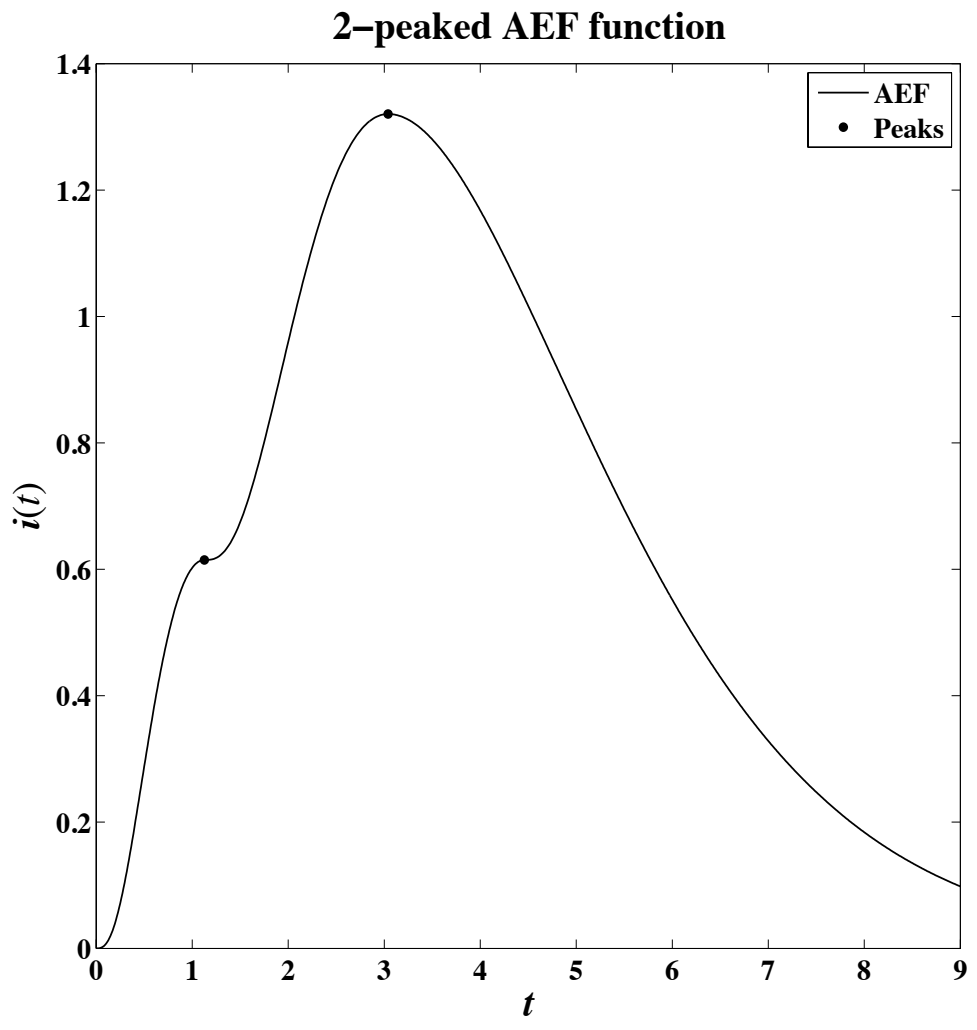

**Fig. 2 –** *An example of a 2-peaked AEF function.*

It this paper, we are exploring the application of the 2-peaked AEF function ($p = 2$ in (4)) to representation of some measured double-peaked current waveforms. Illustration of a 2-peaked AEF function is given in Fig. 2.

## 3 Fitting the *p*-Peaked AEF to Data Using the MLSM

In this section the fitting of the $p$-peaked AEF to some different current waveshapes is explained. The MLSM is used for estimating the $\beta$-parameters, and from these, the corresponding $\eta$–parameters are calculated.

A detailed explanation of the MLSM is not given here, instead we point to [13] for a description of how to apply it in a similar situation. Here, just the parts of this method specific to the use of the $p$-peaked AEF are given.

The MLSM uses a Jacobian matrix that contains the partial derivatives of the residuals and this matrix is denoted by **J**. Suppose that we want to find the least square fit of the $p$-peaked AEF to a set of data points. Then the fitting can be done separately between each peak (and after the final peak). The **J** matrix in this case is

$$\mathbf{J} = \begin{bmatrix} p_{q,1}(t_{q,1}) & p_{q,2}(t_{q,1}) & \cdots & p_{q,n_q}(t_{q,1}) \\ p_{q,1}(t_{q,2}) & p_{q,2}(t_{q,2}) & \cdots & p_{q,n_q}(t_{q,2}) \\ \vdots & \vdots & \ddots & \vdots \\ p_{q,1}(t_{q,k_q}) & p_{q,2}(t_{q,k_q}) & \cdots & p_{q,n_q}(t_{q,k_q}) \end{bmatrix}, \tag{5}$$

where:

- $k_q$ - the number of data points between the $q$th and ($q$-$1$)th peak,
- $t_{q,r}$ - the times corresponding to these data points, and
- $p_{q,r}(t_{q,s}) = \left.\dfrac{\partial i}{\partial \beta_{q,r}}\right|_{t=t_{q,r}}$ more explicitly given by

$$p_{q,r}(t_{q,s}) = 2I_{m_q}\eta_{q,k}\beta_{q,k}h_q(t)x(\beta_{q,k}^2+1), 1 \le q \le p,$$

$$p_{p+1,r}(t_{q,s}) = I_{m_{p+1}}\eta_{p+1,k}\beta_{p+1,k}h_q(t)x(\beta_{p+1,k}^2),$$

where

$$h_q(t) = \ln\left(\frac{t - t_{m_{q-1}}}{\Delta t_{m_q}}\right) + \frac{t - t_{m_{q-1}}}{\Delta t_{m_q}} + 1, 1 \le q \le p,$$

$$h_{p+1}(t) = \ln\left(\frac{t}{t_{m_{p+1}}}\right) + \frac{t}{t_{m_{p+1}}} + 1.$$

The MLSM is an iterative method, and in order to find a new set of $\beta$-parameters in each iteration we also need to find the $\eta$–parameters. This is done by using the regular least square method since for fixed $\beta$-parameters the AEF is linear in $\eta$.

## 4 Numerical Experiments

In this section some results of fitting the 2-peaked AEF function to data given by expressions (6) and (7), Figs. 3 and 4, and also to experimentally obtained ones, Figs. 5 and 6, are presented. For the AEF function the time and current values for the peaks were chosen manually and the rest of the parameters were found using the previously described framework. The number of terms in each interval varies from example to example.

### 4.1 Example 1

Firstly, this method is employed to fit a 2-peaked AEF function to an assumed double-peaked first negative stroke current described in [6], which is expressed by:

$$i(t) = \frac{I_{\max}}{\eta}\left[(1-c)X(t) + cY(t)\right]e^{-t/\tau}, \qquad (6a)$$

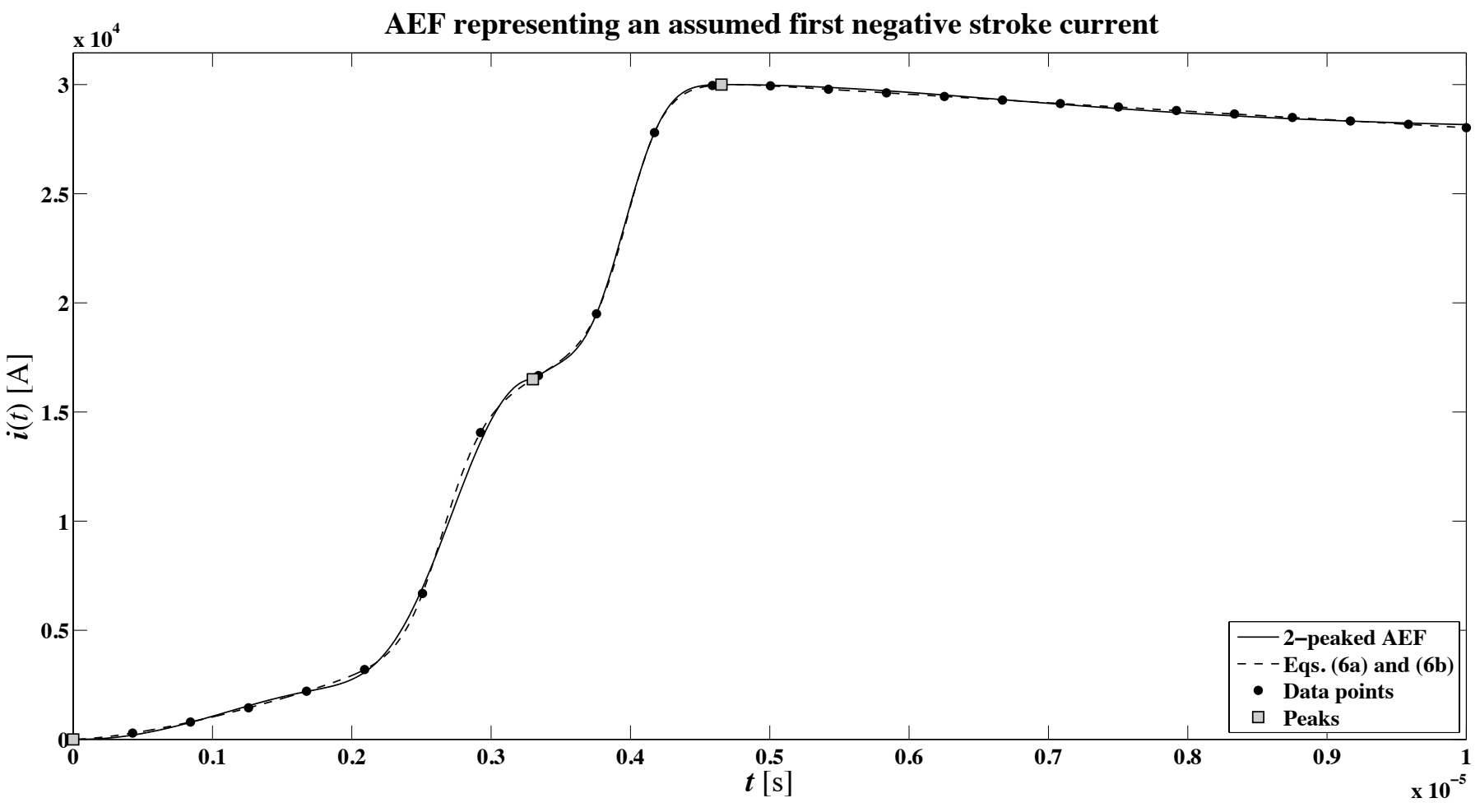

**Fig. 3 –** *The 2-peaked AEF approximating the assumed first negative stroke.*

$$X(t)=\frac{\left(b\frac{t}{T}\right)^{n_2}}{1+\left(b\frac{t}{T}\right)^{n_2}};Y(t)=\frac{a\left(\frac{t}{T}\right)^{k}+\left(\frac{t}{T}\right)^{n_1}}{1+\left(\frac{t}{T}\right)^{n_1}}, \tag{6b}$$

where function parameter values are given in [6], [9]. Figure 3 shows comparison of the fitted AEF function to the one described by (6a) and (6b).

## 4.2 Examples 2 and 3

In these examples, the measured current data for the first negative stroke, acquired at the Mount San Salvatore ([2]), are approximated using the 2-peaked AEF function adjusted by means of the MLSM method. Obtained waveshape in the first 40μs is presented in Fig. 4a, along with the values corresponding to the function proposed in [4] composed of seven Heidler's functions denoted here as MSS_FST#2peaks, which was the goal waveform in the fitting process. It is expressed as:

$$i(t)=\sum_{k=1}^{7}\frac{I_k}{\eta_k}\frac{\left(t/\tau_{1k}\right)^{n_k}}{1+\left(t/\tau_{1k}\right)^{n_k}}e^{-t/\tau_{2k}}, \tag{7}$$

for the parameters listed in [9].

The next example is the approximation of the lightning data recorded at the Morro do Cachimbo station, [19]. Again, the goal waveform was the one proposed in [4] given by (7) with total of twenty-eight adjustable parameters and is here denoted by MCS_FST#2peaks. Illustration of this function, along with the fitted 2-peaked AEF one, is given in Fig. 4b.

In Fig. 4 the quality of the fits varies between the intervals. In the first interval (up to the first peak) the fit is not very good but in the second one the 2-peaked AEF approximates waveshapes well. The AEF may also be used to approximate the peak in the middle of the interval without a need of specifying its position. Since the data points were chosen randomly for this fitting, and the results varied considerably depending on chosen points, some strategy for choosing points should be devised.

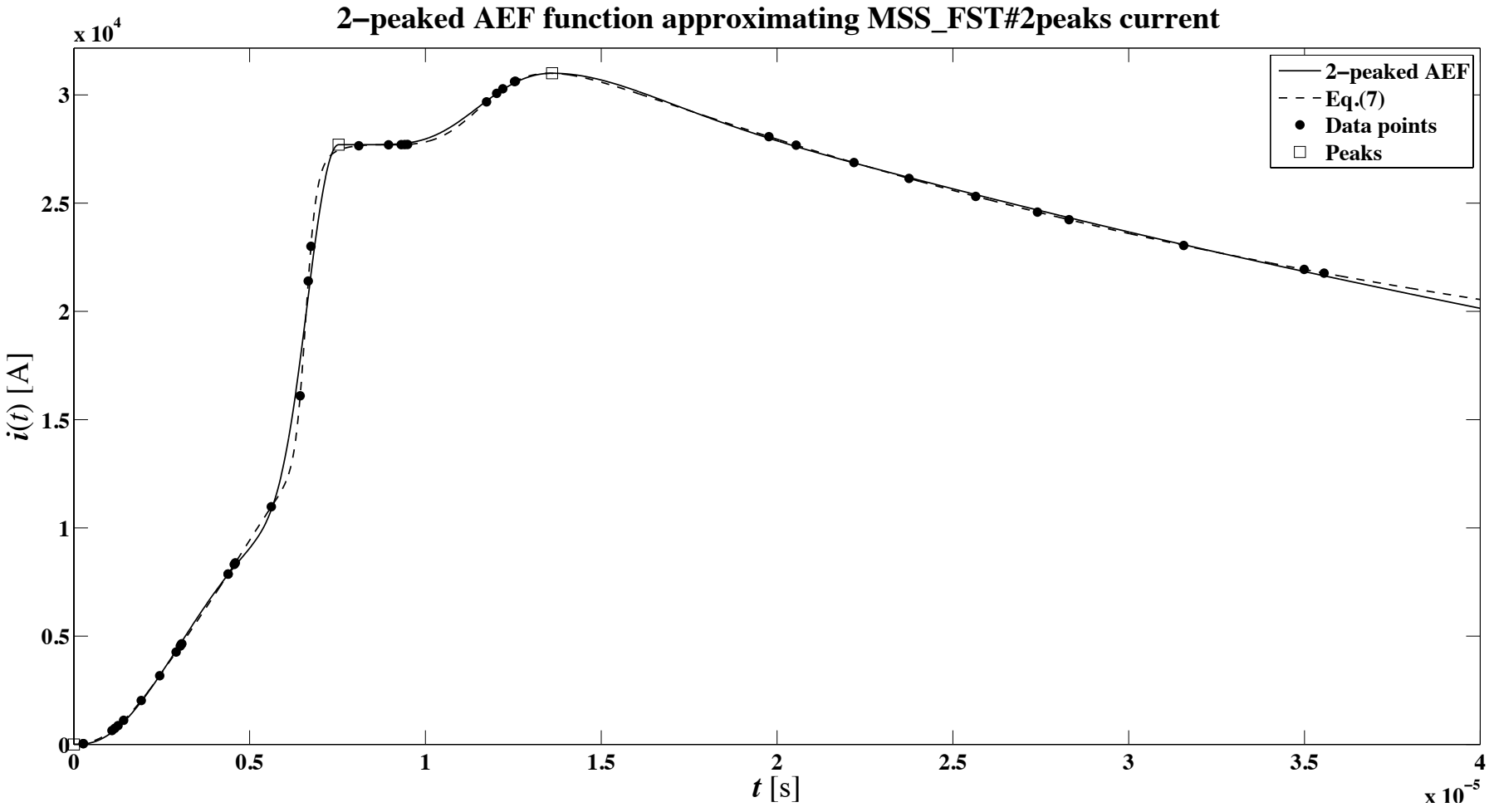

(a)

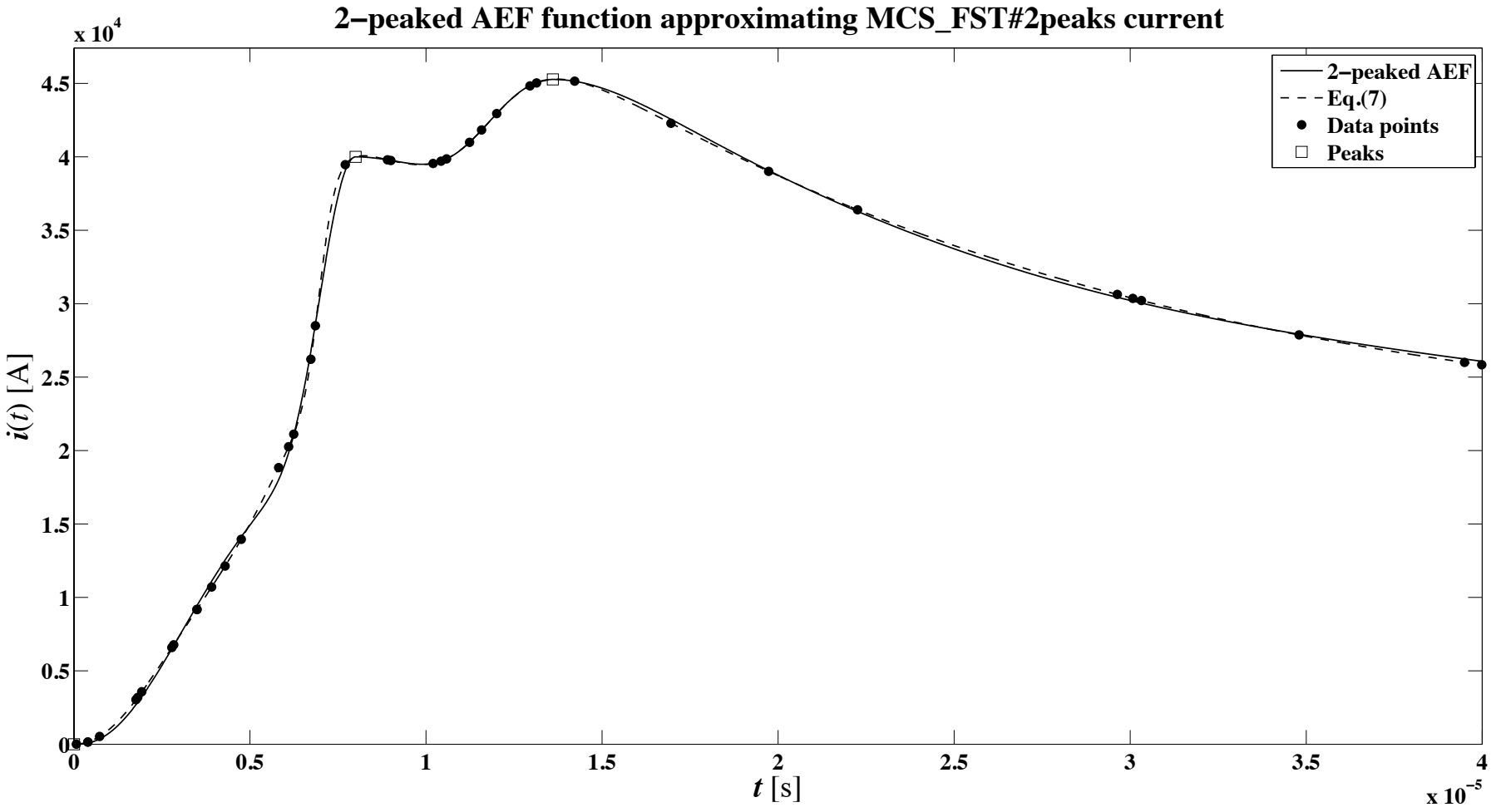

(b)

**Fig. 4 –** *The 2-peaked AEF approximating first negative stroke currents:*
(a) *MSS_FST#2peaks [2].* (b) *MCS_FST#2peaks [19].*

### 4.3 Example 4

Let us now observe another recorded negative first current stroke adopted from [17]. The Heidler's function [11], the Pulse and DEXP functions [13] were optimized using the MLSM in order to be used for representing this set of measured data. The best agreement was achieved using the fitted Heidler's function, however, only in the rising part of the current waveshape. All three functions more or less accurately predict peak current of the measured current, and the time to half-value, but relatively poorly agree with the recorded values in the decaying part. One reason for this probably lays in the double-peaked form of this part of the observed current waveshape.

Visualization of the fitted 2-peaked AEF function is given in Fig. 5. For the sake of comparison, the measured data is also shown. As suggested in [15], the experimentation with the number of terms in each interval could improve the fitting, which was proven right here. This version of AEF fits the given data better than the one presented in [15], with a cost of increased number of terms in the second and third interval.

### 4.4 Example 5

The last example is taken from [3]. In [12], an attempt was made to employ a double exponential function optimized using the MLSM to approximate such current waveshape. However, due to a very steep rise of the DEXP function at $t = 0$, unlike the slow rise of the recorded data, this was not achieved. Certain improvement in the rising part was reached using the Pulse, and even more using a single Heidler function also optimized using the MLSM in [13]. However, the measured waveshape also demonstrates a double-peaked shape of the wavefront, which can only be dealt with more complex functions such as the proposed 2-peaked AEF one.

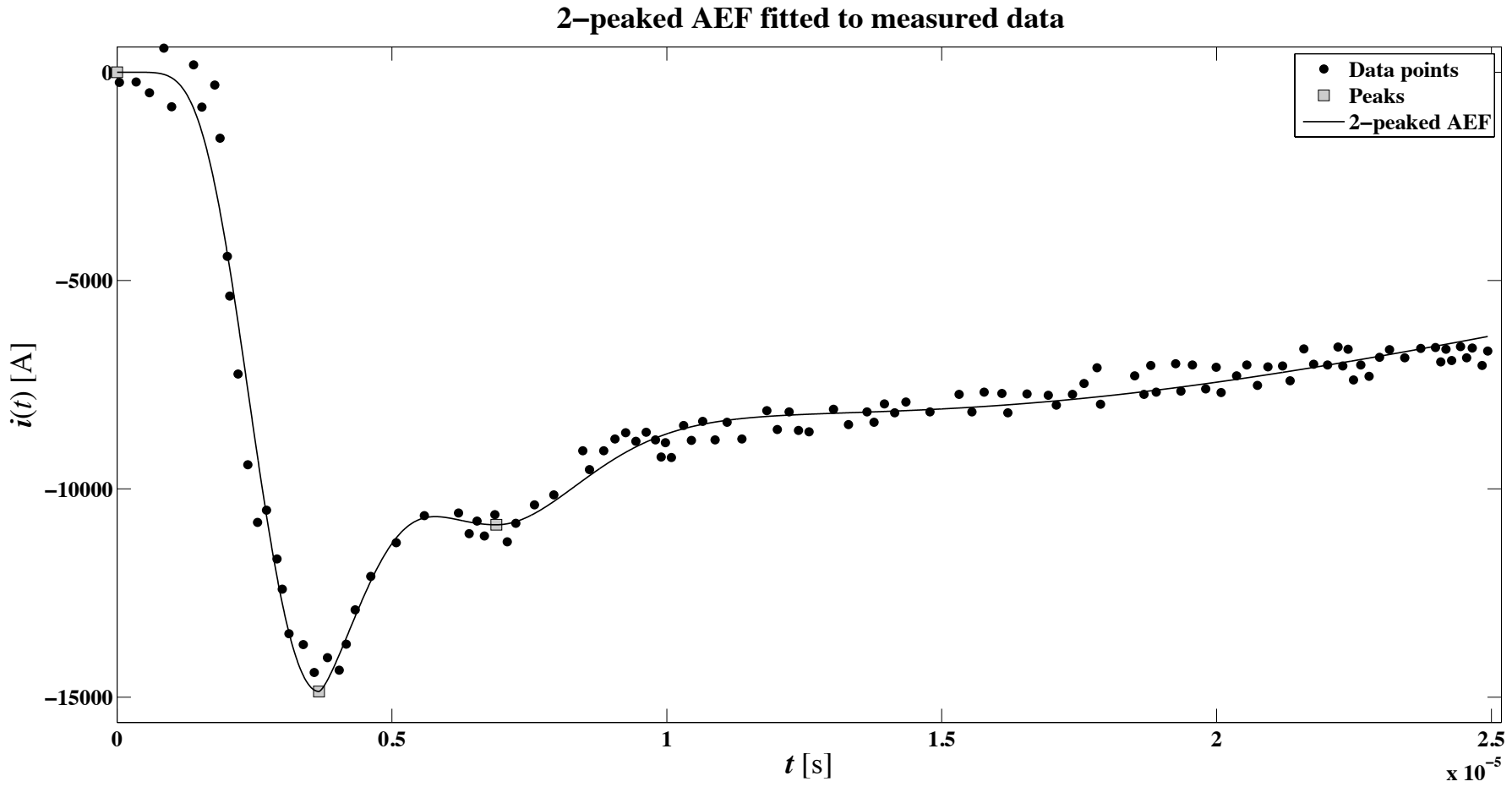

**Fig. 5 –** *The 2-peaked AEF function approximating recorded data from [17].*

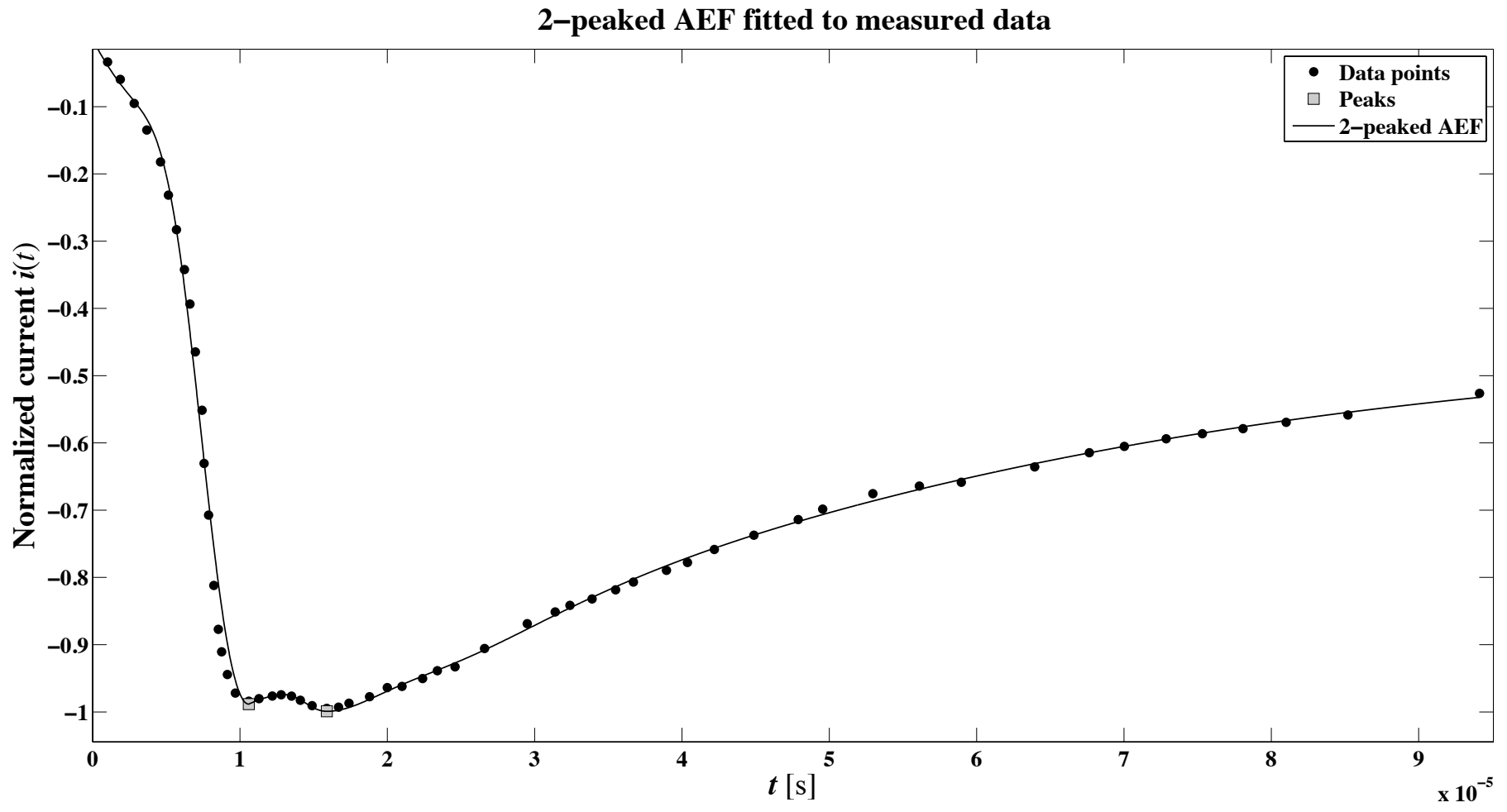

**Fig. 6 –** *The 2-peaked AEF function approximating recorded data from [3].*

Results of performed fitting of the 2-peaked AEF function to this set of recorded data are illustrated in Fig. 6.

## 5 Conclusions

Importance of accurate representation of the first stroke currents is essential in estimating voltages developed by direct strikes and induced voltages by near-by strikes, [18]. The purpose of this work was to employ the 2-peaked AEF function to more effectively represent double-peaked waveshapes typical for experimentally measured first stroke currents, adjusting the function parameters using the MLSM. These are some findings of performed analysis, and possible paths of future work:

- The number of parameters to be optimized is much smaller in comparison to other functions used for this purpose in the literature, [4]. For instance, in Examples 2 and 3 the comparison function had 28 tunable parameters while the AEF had 11; the time and current values for the peaks that were chosen manually (4 of them), and total of 7 exponent values that were tuned using the MLSM (the three AEF intervals had in these examples 2, 2 and 3 terms, respectively).

- The general AEF function (and its 2-peaked version applied here) has analytical expressions for the first derivative, integral, and integral of the square of the function, all necessary for LEMF (lightning electromagnetic field) calculations.

- The $p$-peaked AEF can be used to represent both single-, as shown in [14], [15], and double-peaked current waveshapes as considered here. However, there are some limitations to the waveshapes that can be represented, like that for a chosen (or available) number and distribution of data points and chosen number of terms in the AEF function, the fitting might not be satisfying. This could be in some cases resolved by adding more terms to the linear combination. Also, since in some of the examples the sampling of the model function has been done randomly, which noticeably influenced the final result, a strategy for choosing data points should be found. The authors found that a simple, even distribution of points was unsatisfactory in many cases.

- It should be noted that in all cases MLSM showed some tendencies to find local minima in the objective function instead of the global one, and therefore other methods of fitting should be explored, such as genetic algorithms, trust-region methods, or more specialised methods.

- The first derivative of the *p*-peaked AEF function could be possibly used for approximating desired current derivative waveshapes, which can be of interest when calculating induced effects of lightning, since in certain type of measurements the current derivative is the quantity being measured. This shall be further investigated.